\documentclass[prl,aps,reprint,10pt,amsfonts,amssymb,amsmath]{revtex4-1}

\usepackage{xcolor}
\usepackage{graphicx}
\usepackage{bm}
\usepackage{dsfont}

\renewcommand{\vec}[1]{{\ensuremath{\bm{\mathrm{#1}}}}}

\newcommand{\Tc}{{\ensuremath{T_{\mathrm{C}}}}}

\begin{document}

\title{Static and Dynamical Properties of Antiferromagnetic Skyrmions in the Presence of Applied Current and Temperature}

\author{Joseph Barker}
\affiliation{Institute for Materials Research, Tohoku University, Sendai 980-8577, Japan}

\author{Oleg A. Tretiakov}
\email{olegt@imr.tohoku.ac.jp}
\affiliation{Institute for Materials Research, Tohoku University, Sendai 980-8577, Japan}
\affiliation{School of Natural Sciences, Far Eastern Federal University, Vladivostok 690950, Russia}

\pacs{75.50.Ee 72.25.-b}


\begin{abstract}
Skyrmions are topologically protected entities in magnetic materials which have the potential to be used in spintronics for information storage and processing. However, Skyrmions in ferromagnets have some intrinsic difficulties which must be overcome to use them for spintronic applications, such as the inability to move straight along current. We show that Skyrmions can also be stabilized and manipulated in antiferromagnetic materials. An antiferromagnetic Skyrmion is a compound topological object with a similar but of opposite sign spin texture on each sublattice, which e.g. results in a complete cancelation of the Magnus force. We find that the composite nature of antiferromagnetic Skyrmions gives rise to different dynamical behavior, both due to an applied current and temperature effects. 
\end{abstract}
\maketitle

Skyrmions are topologically protected objects which can form in magnetic materials~\cite{Nagaosa:2013cc}. They are an active area of research in spintronics because of their potential for encoding, transmitting~\cite{Sampaio:2013wc,Fert:2013ul} and computing information~\cite{Zhang:2015cs}. Belavin and Polyakov~\cite{Polyakov75} introduced Skyrmions \cite{Skyrme1962}  in the context of the two-dimensional Heisenberg model although Feldtkeller~\cite{Feldtkeller65} and Thiele~\cite{Thiele73} discussed similar topological defects earlier in a more general context.
Skyrmions have some advantages over other proposed storage technologies, such as domain wall registers, because of the low currents required to move them due to the ability to move past pinning sites~\cite{Iwasaki2013}. Implementing Skyrmion devices in ferromagnetic materials involves issues in common with other spintronic concepts, such as the sensitivity to stray fields. Skyrmions also posses a further complication in that they experience a Magnus force perpendicular to the applied current, making it difficult to move Skyrmions along the current~\cite{Iwasaki:2013ji}.
By contrast, antiferromagnets are not sensitive to stray fields and with an applied current we find that Skyrmions in antiferromagnets move in straight lines along the current, distinctly different from ferromagnetic materials. In an antiferromagnet the Skyrmion forms as a pair of strongly coupled topological objects, one pertaining to each sublattice. The opposing topological index of each sublattice causes an exact cancellation of the Magnus force, hence there is no transverse component of the velocity. The current induced longitudinal velocity is also found to strongly depend on the material parameters ($\alpha$, $\beta$) and as a result can reach high velocities of the order of km/s. Moreover, the thermal properties of  antiferromagnetic Skyrmions are found to be rather different from their ferromagnetic counterparts.

A compelling reason to study Skyrmions in antiferromagnets apart from the insensitivity to stray fields, is that the Dzyaloshinskii-Moriya interaction (DMI), which is essential for the formation of individual Skyrmions, is more commonly found in antiferromagnetic (AFM) materials than ferromagnetic (FM) materials. Most recent experimental results on FM Skyrmions rely on the presence of an interfacial DMI to stabilize Skyrmions, however bulk DMI is more prevalent in AFMs~\cite{Dzyaloshinskii58, Moriya60}. AFMs are also considerably more abundant in nature than ferromagnets, although metallic AFMs are not so common but examples include FePt$_3$ and Mn$_2$Au.

Skyrmions can form in different systems where there is a competition between the DMI and another energy contribution, for example the Zeeman energy from an applied field or a uniaxial anisotropy~\cite{Fert:2013ul,Woo:2015vl}. Here the last option is studied by necessity because of the antiferromagnets insensitivity to applied fields and the lack of a significant demagnetizing field precludes these mechanisms from forming a Skyrmion \cite{Bogdanov2002}. We also focus on individual Skyrmions, rather than a Skyrmion lattice, as the ability to move and manipulate individual bits of information is more relevant to the suggested technological applications \cite{Sampaio:2013wc}.

\begin{figure}
    \includegraphics[width=\linewidth]{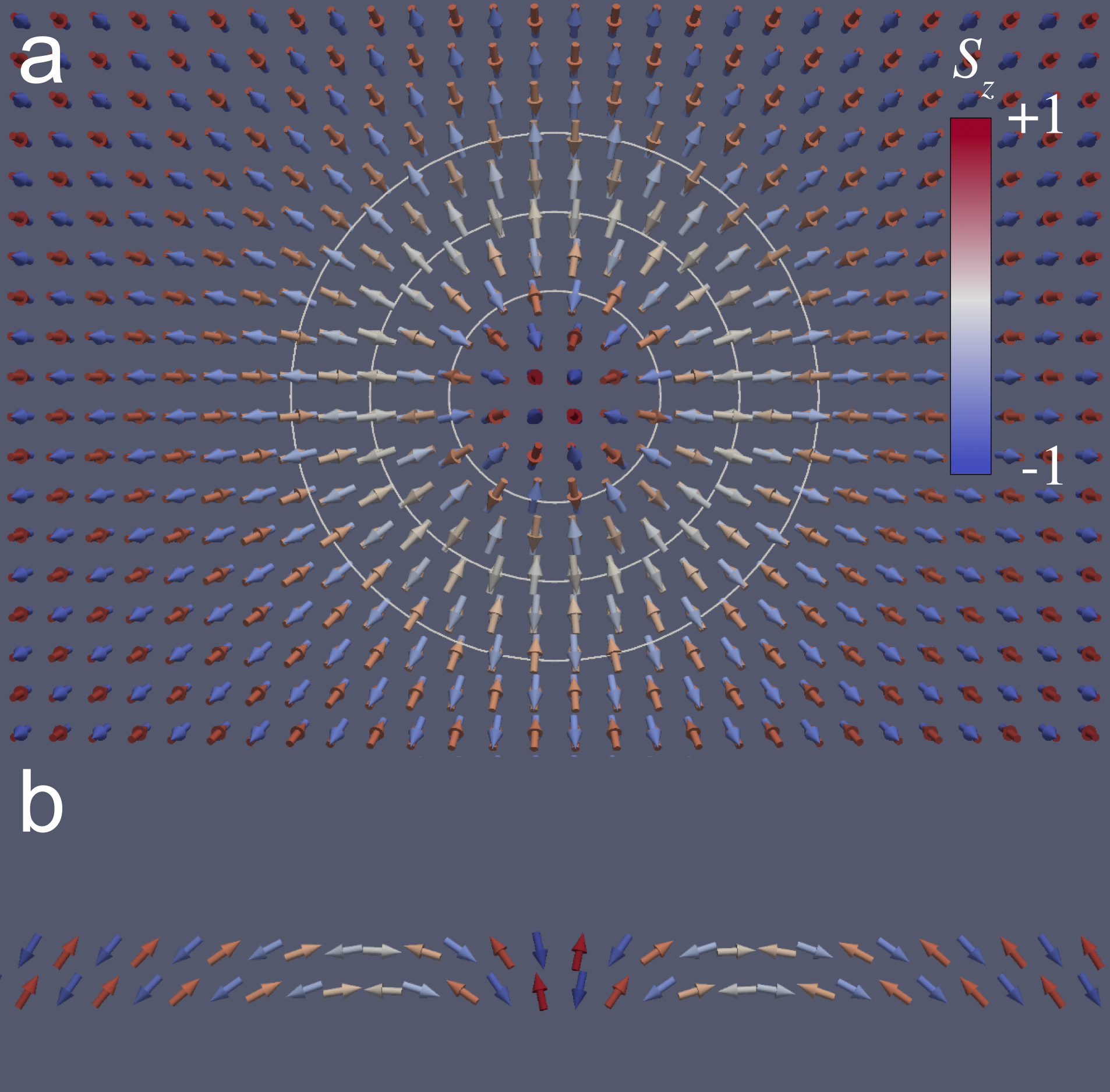}
    \caption{
        The spin texture of G-type AFM Skyrmion. a) top view of the Skyrmion, white lines show contours of constant $n_z$. The radius is $2.1$~nm. b) cross-section of the Skyrmion. The core is not a single spin but a compensated structure combining the two sublattices.
        \label{fig:af_Skyrmion_montage}
    }
\end{figure}

\begin{figure}
    \includegraphics[width=\linewidth]
        {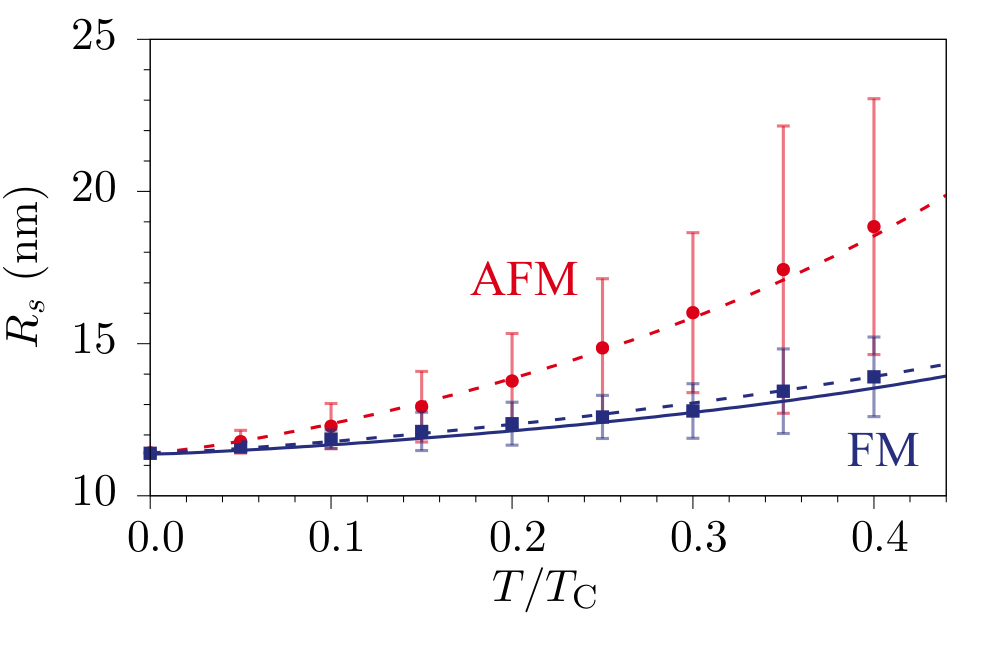}
    \caption{
        \label{fig:af_Skyrmion_radius}
        AFM and FM Skyrmion radii $R_s$ as a function of reduced temperature $T/\Tc$. The error bars represent the magnitude of the thermally induced radius fluctuations. The solid line is the simple scaling theory, Eq.~(\ref{eq:Skyrmion_radius}), dashed lines are power law fits $R_s(T) \propto m^{-0.86}$ and $R_s(T) \propto n^{-2.09}$.
    }
\end{figure}

In this Letter we consider the so-called `G-type' antiferromagnet, formed by a three-dimensional chess board like pattern. The AFM Skyrmion forms in much the same way as a FM Skyrmion, by introducing a topological defect, reversing the $A$ and $B$ sublattices within a small area and allowing the system to relax. The DMI prevents the metastable domain from reversing. The spin structure shown in Fig.~\ref{fig:af_Skyrmion_montage} is analogous to the `hedgehog' Skyrmion state of a FM but with one of the sublattices inverted. Hence, the topological defect exists in the N\'{e}el field and the magnetization is nearly zero everywhere. At the center of the AFM Skyrmion, neither sublattice dominates, but instead there is a compensation of opposite spins around the true center of the Skyrmion. We find that in the absence of temperature the radial profile and Skyrmion radius for a given DMI are the same in both FM and AFM Skyrmions for the magnetization and N\'{e}el parameters respectively (Fig.~\ref{fig:af_Skyrmion_radius}).

We first study the athermal dynamics of the AFM Skyrmion with an applied current, comparing the fundamentals of AFM Skyrmion dynamics with Thiele's equations \cite{Thiele73, Tretiakov08, Clarke08, Everschor2011, Iwasaki2013} for FM spin textures, before moving onto more complicated effects introduced by temperature. Coupling to the current assumes that the electrons of up and down spin are transported predominantly through their corresponding magnetization sublattice~\cite{Cheng:2014gc}. For the G-type AFM this is reasonable, but for other AFMs the transport of the electronic current through the AFM may be different. From this assumption the spatial derivative $\vec{\nabla} \vec{M}$ is calculated for the magnetization of each sublattice, rather than the net local magnetization which is almost zero.

\begin{figure}
    \includegraphics[width=\linewidth]
        {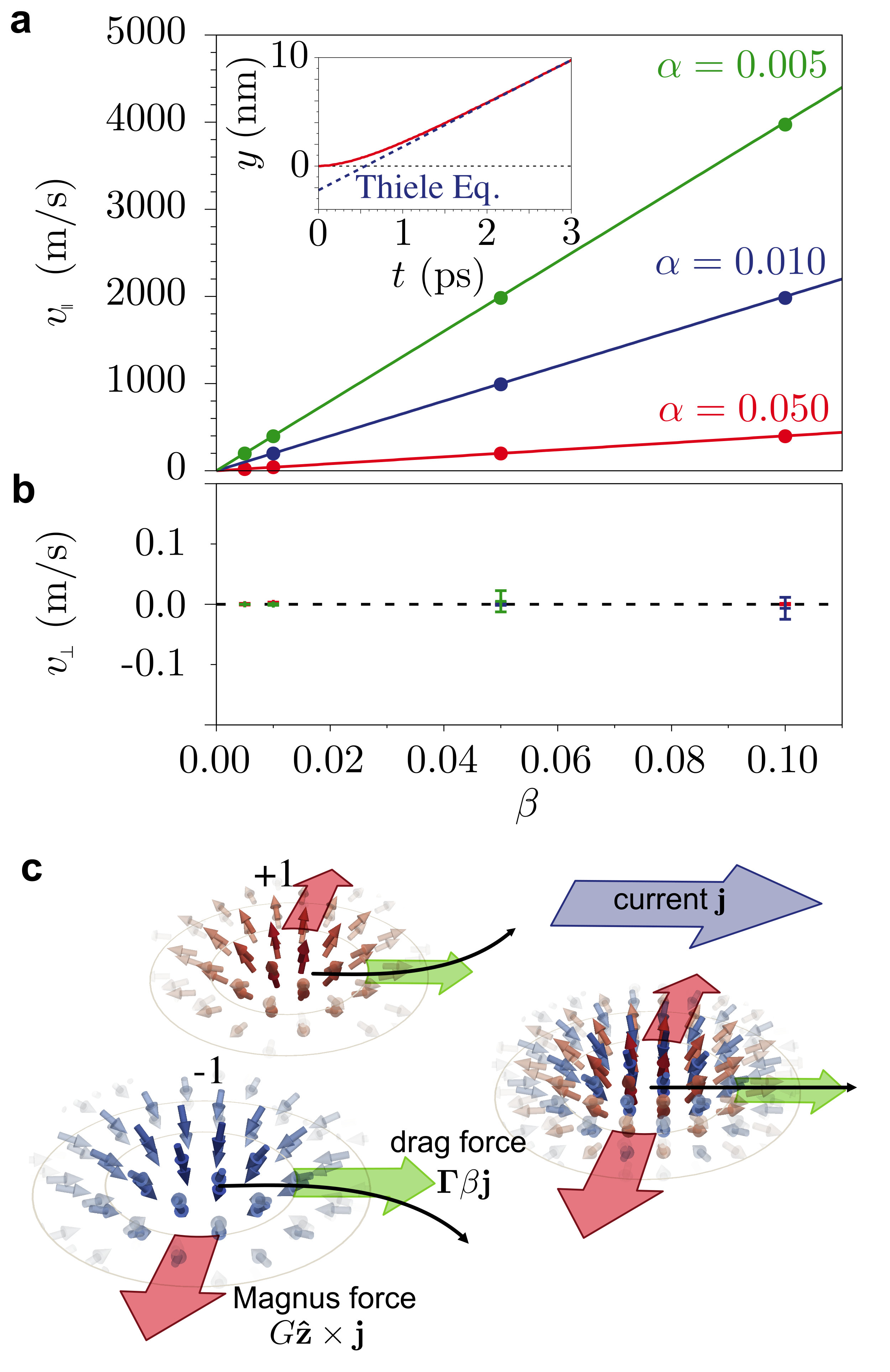}
    \caption{
        \label{fig:af_Skyrmion_velocity}
        Current induced AFM Skyrmion dynamics ($j= 200$ m/s). a) longitudinal velocity for different combinations of $\alpha$ and $\beta$. Points are calculated numerically and the lines are Eq.~(\ref{eq:thiele_parallel}) based on Thiele's equations for an AFM. Inset shows the mass term is small and the Skyrmion reaches terminal velocity after $2$~ps. b) transverse velocities calculated from the same simulations show there is no transverse motion. c) the AFM Skyrmion is composed of two topological objects with opposite topological charge, hence the Magnus force acts in opposite directions. The strong coupling between the sublattices leads to a perfect cancellation of the two opposing forces and so the AFM Skyrmion has no transverse motion.
    }
\end{figure}

Comparing the AFM dynamics with those of a FM Skyrmion (where the only change in material parameters is in the sign of the exchange interaction), highlights two main intrinsic differences in the dynamics resulting from the AFM characteristics. Firstly, the AFM Skyrmion always has zero transverse velocity $v_{\perp}$, relative to the current. In the FM this is only true for the highly symmetric case of $\alpha = \beta$, where $\alpha $ is the Gilbert damping constant and $\beta$ is the non-adiabatic spin-transfer torque parameter. In a FM the transverse velocity is due to the Magnus force acting on the Skyrmion and the direction ($\pm \hat{y}$) is determined by the winding number of the Skyrmion:
\begin{equation}
    \label{eq:skyr_number}
        Q^{(k)} = \int \frac{d^2r}{8\pi} \, \epsilon_{ij} \, \epsilon_{\alpha\beta\gamma}
    \, m_\alpha^{(k)} \, \partial_i m_\beta^{(k)} \, \partial_j m_\gamma^{(k)}, 
\end{equation}
where $\vec{m}^{(k)}(\vec{r})$ is the unit vector parallel to the
local magnetisation $\vec{M}^{(k)}(\vec{r})$ and $k=1,2$ label the sublattices in the AFM case. The AFM Skyrmion is essentially composed of two topological objects with opposite winding numbers ($Q^{(k)} =\pm 1$) which are strongly coupled through the AFM exchange interaction. Both sublattices generate a Magnus force, but there is a perfect cancellation (Fig.~\ref{fig:af_Skyrmion_velocity}c) thus resulting in no $v_{\perp}$. As a result the AFM Skyrmion travels in a perfectly straight trajectory along the current (see Supplementary Movie S1). One can also directly define the winding number for the AFM order parameter (N\'{e}el field) $\vec{n}(\vec{r},t)=\mathbf{m}^{(1)}(\vec{r},t)-\vec{m}^{(2)}(\vec{r},t)$ and thus show that AFM Skyrmions are topologically nontrivial textures with the AFM topological charge $\pm 1$. 

The second notable difference between AFM and FM Skyrmions is that the longitudinal velocity in the AFM can greatly exceed the FM Skyrmion drift velocity which is always close to the electron drift velocity ($v \approx 200$ m/s for current $j= 200$ m/s and $\alpha, \beta \ll 1$). For low $\alpha$ or high $\beta$, AFM Skyrmions can move at km/s whilst remaining stable (Fig.~\ref{fig:af_Skyrmion_velocity}a). Recent theoretical studies of AFM dynamics give an insight into this~\cite{Tveten2013, Cheng2014, Kim2014}. The dynamics can be studied based on the generalized Thiele's equations~\cite{Tveten2013} which describe the motion of a spin texture in terms of collective coordinates $b_j$ 
\begin{equation}
\label{eq:eqmotion}
    \mathcal{M}^{ij}\ddot{b}_{j}+ \alpha\Gamma^{ij} \dot{b}_{j} = F^{i},
\end{equation}
which correspond to the soft modes of the Skyrmion~\cite{sup_material}. Here $\mathcal{M}^{ij}$ is the mass tensor, $\alpha\Gamma^{ij}$ characterizes viscous friction and is related to damping in the AFM, and finally $F^{i}$ is the generalized force due to the current. The mass term contributes to the dynamics only on short time scales (up to $\sim 2$ ps), as shown in the inset of Fig.~\ref{fig:af_Skyrmion_velocity}a. The AFM Skyrmion reaches its terminal velocity quickly, which is understood more intuitively by switching back to the two-sublattice description and writing the Thiele's equations for $\mathbf{m}^{(1,2)}(\mathbf{r},t)$~\cite{sup_material}. It is then clear that the Magnus forces for $\mathbf{m}^{(1)}(\mathbf{r},t)$ and $\mathbf{m}^{(2)}(\mathbf{r},t)$, $G\hat{\mathbf{z}}\times \mathbf{j}$ with the gyrocoupling constant $G=4\pi Q^{(k)}$, cancel each other (Fig.~\ref{fig:af_Skyrmion_velocity}c) and the remaining generalized drag force, $F=\Gamma \beta j$, leads to the AFM Skyrmion velocity
\begin{equation}\label{eq:thiele_parallel}
    v_{\parallel} = \frac{\beta}{\alpha} j
\end{equation}
which is only along the direction of the current. The velocity is plotted as lines in Fig.~\ref{fig:af_Skyrmion_velocity}a, showing an excellent agreement with the simulations. 

The importance of understanding the thermal properties of Skyrmions is now becoming clear~\cite{Schutte2014}. On a macroscopic level Skyrmions diffuse due to the thermal perturbations of the magnetic moments. We also consider the temperature dependence of the macroscopic material parameters, such as anisotropy and exchange stiffness, and find this can lead to a change in the balance of the competing energy terms in Skyrmionic systems. The Skyrmion spin texture is also subjected to deformations due to internal dynamics which are stimulated by thermally induced spin waves. Using Langevin Landau-Lifshitz-Gilbert simulations~\cite{sup_material} we have studied these thermal effects for the AFM Skyrmion and compared it with that of the FM Skyrmion (see Supplementary Movies S3 and S4).

\begin{figure}
    \includegraphics[width=\linewidth]
        {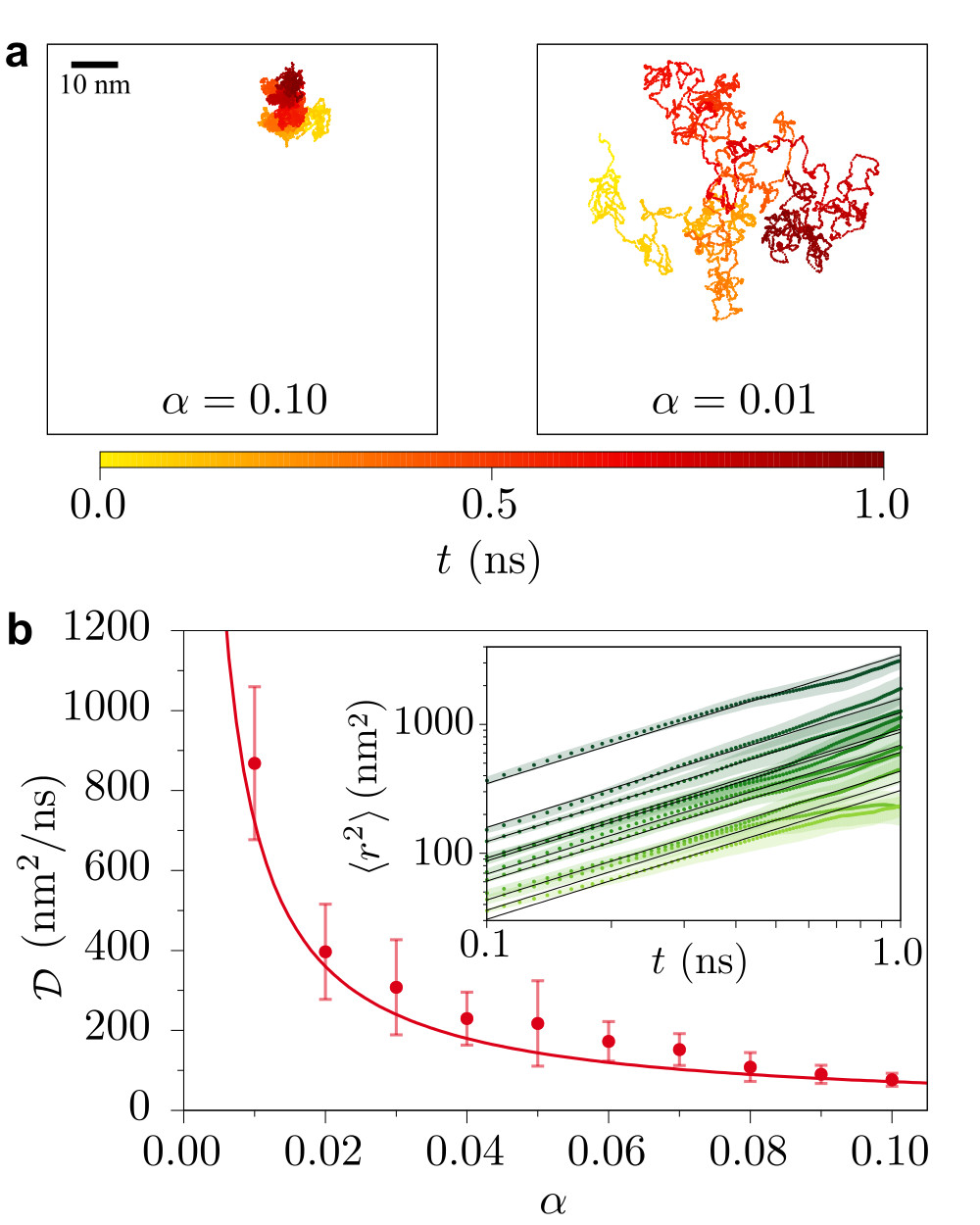}
    \caption{
        \label{fig:af_Skyrmion_diffusion}
        a) Brownian motion of the AFM Skyrmion for Gilbert damping constants $\alpha = 0.1$ and $0.01$ at $T/T_c = 0.25$. b) Diffusion coefficient of the AFM Skyrmion as a function of $\alpha$. Points are calculated from the mean squared displacement $\langle r^2 \rangle$ (inset), the solid line is $\mathcal{D} = \lambda k_B T/ (2\alpha s \sigma)$.
    }
\end{figure}

Simulating the Brownian motion of a single Skyrmion (Fig.~\ref{fig:af_Skyrmion_diffusion}), we find the AFM Skyrmion to be diffusive, meaning the mean square displacement $\langle r^{2} \rangle\propto t$, as was shown for the FM Skyrmion~\cite{Schutte2014}. However the diffusion coefficient of the AFM Skyrmion is greater than that of the FM and $D \propto 1/\alpha$ as a direct consequence of the absence of a net Magnus force which plays the dominant role in the FM diffusion where $D\propto \alpha$~\cite{Schutte2014}. In Ref.~\onlinecite{Kim2015} the diffusion coefficient of AFM textures in one-dimensional case was shown to be $\mathcal{D} = \lambda k_B T/ (2\alpha s \sigma)$, i.e. inversely proportional to the Gilbert damping $\alpha$, where $s=\hbar S / a^{3}$, $\lambda=\sqrt{A/K}$ is the exchange length and $\sigma$ is the cross sectional area of the domain wall - in this case the circumferential area of the Skyrmion, $2\pi a R_{s}(T)$. The lack of a Magnus force allows this expression to be generalized to two-dimensional diffusion, we compared it with our numerical results (solid line in Fig.~\ref{fig:af_Skyrmion_diffusion}, no fitting was performed) finding a good agreement.

The increased thermal mobility of the AFM Skyrmions may be useful if attempting to move spin textures with heat gradients. However, it also poses challenges for more conventional current driven motion as the thermal perturbations may cause excessive randomness in the motion. We have studied just one of the AFM types in this Letter and it is thus possible that other AFMs could contain the benefit of high drift velocity but with a lower diffusion coefficient.

We also calculated the temperature dependence of the mean Skyrmion radius, $R_s$, and compared to an estimate from the scaling relationships of $A$, $D$ and $K$ with the reduced magnetization $m = M(T)/M(0)$. The scaling of $A \propto m^{2}$ and $K \propto m^{3}$ in bulk FMs are well known and we use these as approximations in the thin film. In the absence of knowledge concerning the thermal scaling of $D$ we find that assuming no temperature dependence of $D$ gives the closest agreement with the numerical results (Fig.~\ref{fig:af_Skyrmion_radius}), the final equation being
\begin{equation}
\label{eq:Skyrmion_radius}
    R_{s}(T) = \sqrt{\frac{2A\lambda}{4\sqrt{AK}m(T) - \pi Dm^{-3/2}(T)}} .
\end{equation}

The results show that increasing temperature causes a change in the domain wall energy cost, with the effective DMI contribution becoming larger as temperature increases, causing an increase in the Skyrmion radius \footnote{This result remains true within the error bars even with a temperature dependence of $D \propto m^2$.}. Surprisingly the AFM Skyrmion radius shows a stronger temperature dependence and larger fluctuations of the radius about the mean value than for the FM Skyrmion. At zero temperature we found the FM and AFM Skyrmion to have the same $R_s$ vs. $D$ dependence as shown by Rohart \textit{et al}.~\cite{Rohart:2013ef} for FM Skyrmions. Hence the deviation observed at finite temperature is the result of the different characteristic thermal fluctuations in an FM and AFM. 

The plethora of different types of antiferromagnetic materials makes further research in the area of AFM Skyrmions an exciting prospect. Searching for AFM Skyrmions experimentally will be a challenging task, but techniques such as neutron scattering have the potential to find AFM Skyrmion lattices. Moreover, x-ray magnetic linear dichroism occurs in the AFM state as the spin-orbit coupling leads to a distortion of the charge density and, thus the AFM structure can be experimentally measured \cite{GLaan2013}.

Small spin textures can serve as bits of information and manipulating them by electric or thermal currents one of the main challenges in the field of spintronics. Ferromagnetic Skyrmions recently attracted a lot of attention because of their small size and ability to avoid pinning while moved by electric current better than domain walls. However, ferromagnetic Skyrmions still suffer from the detrimental effects of stray fields and transverse intrinsic dynamics causing difficulties in employing them in spintronic applications. The related topological object we explored here -- the AFM Skyrmion -- overcomes these disadvantages, having no demagnetizing field, an insensitivity to stray fields, and we have shown that its dynamics are strictly along the current (no Hall effect) while potentially being faster compared to its ferromagnetic analogue. This makes AFM Skyrmions an ideal information carrier. However, the thermal properties, such as the AFM Skyrmion radius and diffusion constant differ from those for ferromagnetic Skyrmions because of the fundamental difference in the spin correlations. More studies into the different AFM types and the effect of the spin correlations are needed to discover if the benefits of the AFM Skyrmions can be achieved but with a reduced diffusivity than we have found here.

The authors thank M. Kl\"{a}ui, J. Sinova, G. Tatara, and Y. Tserkovnyak 
for helpful discussions. O.A.T. and J.B. acknowledge support
by the Grants-in-Aid for Scientific Research (Grants No. 25800184,
No. 25247056, No. 25220910, and No. 15H01009) from the Ministry of Education, Culture, Sports, Science and Technology
(MEXT) of Japan and SpinNet. O.A.T. is indebted to KITP at UC, Santa Barbara for hospitality, this research was supported in part by the National Science Foundation under Grant No. NSF PHY11-25915. J.B. acknowledges support from the Graduate Program in Spintronics, Tohoku University.

\textit{Note added in proof.} --
Recently, we became aware of a recent report \cite{Zhang2015} on another study of AFM Skyrmions.

%


\end{document}